\documentclass[aps,prb,twocolumn,superscriptaddress,floatfix]{revtex4-1}
\usepackage{graphicx}
\usepackage{amsmath}
\usepackage{amsthm}
\usepackage{dsfont}
\usepackage{xcolor}

\def\Aarhus{Department of Physics and Astronomy, Aarhus University, Ny Munkegade 120, DK-8000 Aarhus C}

\def\PKS{Max-Planck-Institut f\"{u}r Physik komplexer Systeme, D-01187 Dresden, Germany}

\begin{document}

\title{Truncation of lattice fractional quantum Hall Hamiltonians\\ derived from conformal field theory}

\author{Dillip K. Nandy}
\thanks{These authors contributed equally to this work.}
\affiliation{\Aarhus}

\author{N. S. Srivatsa}
\thanks{These authors contributed equally to this work.}
\affiliation{\PKS}

\author{Anne E. B. Nielsen}
\altaffiliation{On leave from \Aarhus}
\affiliation{\PKS}

\begin{abstract}
Conformal field theory has recently been applied to derive few-body Hamiltonians whose ground states are lattice versions of fractional quantum Hall states. The exact lattice models involve interactions over long distances, which is difficult to realize in experiments. It seems, however, that such long-range interactions should not be necessary, as the correlations decay exponentially in the bulk. This poses the question, whether the Hamiltonians can be truncated to contain only local interactions without changing the physics of the ground state. Previous studies have in a couple of cases with particularly much symmetry obtained such local Hamiltonians by keeping only a few local terms and numerically optimizing the coefficients. Here, we investigate a different strategy to construct truncated Hamiltonians, which does not rely on optimization, and which can be applied independent of the choice of lattice. We test the approach on two models with bosonic Laughlin-like ground states with filling factor $1/2$ and $1/4$, respectively. We first investigate how the coupling strengths in the exact Hamiltonians depend on distance, and then we study the truncated models. For the case of $1/2$ filling, we find that the truncated model with truncation radius $\sqrt{2}$ lattice constants on the square lattice and $1$ lattice constant on the triangular lattice has an approximate twofold ground state degeneracy on the torus, and the overlap per site between these states and the states constructed from conformal field theory is higher than $0.99$ for the lattices considered. For the model at $1/4$ filling, our results give some hints that a truncation radius of $\sqrt{5}$ on the square lattice and $\sqrt{7}$ on the triangular lattice might be enough, but the finite size effects are too large to judge whether the topology is, indeed, present in the thermodynamic limit. The states with high overlap also have the expected topological entanglement entropies.
\end{abstract}

\maketitle

\section{Introduction}
The fractional quantum Hall (FQH) effect is the precursor to a wide variety of systems that exhibit topological order and showed the limitations of Landau's theory of symmetry breaking. Among several attempts to comprehend the phenomenon, Laughlin's intuitive ansatz\cite{laughlin} for the wavefunction successfully captured the physics to a great extent. In addition to studying the FQH effect in continuum systems, it is also interesting to realize FQH physics on lattices.\cite{Hormozi, Hafezi, Goldman, Palmer} In lattices, the physical magnetic field can be replaced by an artificial magnetic field, and a larger gap in the spectrum can be achieved. Another strong motivation for studying lattices is the interest in realizing FQH physics in ultracold atoms in optical lattices, which would pave the way for very detailed experimental investigations of the effect. In addition, lattice systems are known to host qualitatively new phases of matter.\cite{Emil, Yang}

One route to obtain FQH physics in lattices is to engineer flat bands that bear nonzero Chern numbers. The bands mimic Landau levels, and FQH like states can be achieved by adding interactions.\cite{titus, cherna, chernb} Another route is to construct lattice versions of FQH trial wavefunctions\cite{KL} and derive parent Hamiltonians for these states.\cite{Schroeter,Thomale,kapitz,Nielsen} The Hamiltonians obtained this way can be different from the fractional Chern insulator Hamiltonians.

It has turned out\cite{MooreRead} that a number of analytical FQH trial wavefunctions can be expressed in terms of conformal field theory (CFT) correlation functions, and a closely related construction can be used for lattice states. Using null fields in the CFT provides a route to construct parent Hamiltonians for the lattice states.\cite{Nielsen} The CFT approach to derive exact lattice Hamiltonians begins by finding a set of operators $\Lambda_i$ which annihilate the lattice FQH state $|\Psi_{\textrm{FQH}}\rangle$, i.e.\
\begin{eqnarray}
\Lambda_i & |\Psi_{\textrm{FQH}} \rangle = 0.
\end{eqnarray}
The Hamiltonian, whose ground state is $|\Psi_{\textrm{FQH}}\rangle$, is then given by the following positive semi-definite operator
\begin{eqnarray}
H^{\textrm{Exact}} = \sum_{i} \Lambda^{\dagger}_i \Lambda_i. \label{ham}
\end{eqnarray}
Typically, this Hamiltonian consists of few-body interactions and is nonlocal.

There are more reasons why it would be desirable to have a local Hamiltonian rather than a nonlocal Hamiltonian. First, long-range interactions are difficult to achieve in experiments. Second, it would be natural for the terms of a local Hamiltonian to only depend on the local structure of the lattice and not on all sites in the whole lattice, which also naturally leads to a well-defined thermodynamic limit. This simplifies the description and the interpretation. Third, the local Hamiltonians are easier to work with numerically, e.g.\ when studying the systems with exact diagonalization or tensor networks.\cite{chen}

The FQH states have correlations that decay exponentially with distance in the bulk. This suggests that it should, in fact, not be necessary to have long-range interactions in the Hamiltonians. In previous work,\cite{natureanne,glasser} local Hamiltonians for a bosonic Laughlin state and a bosonic Moore-Read state with SU(2) symmetry were achieved using the following procedure. First, all terms except a selection of local terms are removed from the Hamiltonian. The coefficients of the remaining terms are then adjusted so that all terms that only differ by a lattice translation have the same strengths. Finally, the relative strengths of terms that do not only differ by a lattice translation are determined by a numerical optimization of the overlap between the CFT states and the ground states of the truncated models. The latter step involves exact diagonalization and is done for small system sizes. This way of truncating the Hamiltonian is easier for models that have SU(2) symmetry, since the symmetry reduces the number of possible terms in the Hamiltonian and hence also the number of coefficients that need to be optimized. In models that do not possess such symmetries, however, many more terms in the Hamiltonian are possible, and this makes it difficult to find a suitable choice of parameters, which gives a high overlap per site independent of the system size.\cite{glasser}

In this article, we investigate a different procedure to truncate Hamiltonians derived from CFT. The procedure is more general and can be applied to Hamiltonians with or without symmetries on all lattice geometries. The idea is to truncate the operator $\Lambda_i$ in the first place and then proceed to construct the Hamiltonian as in Eq.\ \eqref{ham}. There are more advantages of this approach. The $\Lambda_i$ operator is simpler to truncate than the Hamiltonian, since it contains fewer terms, and since the coefficients of the terms depend only on the relative positions of the involved sites and not on the rest of the lattice. This means that the local terms of the resulting Hamiltonians are independent of the size of the lattice used to compute them. Another advantage is that it is clear how to obtain Hamiltonians with either periodic or open boundary conditions as desired. Finally, there is no optimization involved, and once the lattice has been chosen, the only input to the procedure is the truncation radius for the $\Lambda_i$ operator.

We test the proposed truncation procedure numerically for half filled and quarter filled bosonic Laughlin states on square and triangular lattices. We do this by computing the overlap between the analytical states and the ground states of the truncated Hamiltonians, and by computing the topological entanglement entropy. For the half filled model on the torus, we find that there is a near twofold degeneracy with a gap to the first excited state as one would expect as a consequence of the topological order. The overlap per site is higher than $0.99$ for the two lowest energy states for all cases considered. For the quarter filled model, we find that a larger truncation radius is needed to get high overlaps for low energy states on small lattices.

The paper is structured as follows. In Sec.\ \ref{sec:model}, we introduce the exact models that we use for testing the truncation approach. We also show that truncating the Hamiltonian directly does not give an unequivocal result. In Sec.\ \ref{sec:local}, we derive the local models. In Sec.\ \ref{sec:gsprop}, we report our numerical results for the overlap between the analytical states and eigenstates of the truncated models. We also investigate the ground state degeneracies. Section \ref{sec:conclusion} concludes the paper, and the appendix provides expressions for the exact wavefunctions on the torus.

\section{Exact Model}\label{sec:model}
\subsection{Exact Wavefunction}\label{sec:exwf}
Let us first introduce the exact form of the considered wavefunctions on a lattice on the plane (i.e.\ with open boundary conditions) before constructing the exact Hamiltonians. The lattice Laughlin state with one particle per $q$ flux lines, where $q$ is a positive integer, can be expressed as\cite{TuNJP}
\begin{equation}
 |\Psi_{\textrm{Exact}}^{\textrm{P}}\rangle=
 \sum_{n_1,n_2,\ldots,n_N}
 \Psi(n_1, n_2, \ldots, n_N)
 |n_1,n_2,\ldots,n_N\rangle  \label{state}
 \end{equation}
with
 \begin{eqnarray}
 \Psi(n_1, n_2, \ldots, n_N)
 \propto \delta_n \chi_n \prod_{i<j}
 (z_i - z_j)^{qn_in_j-n_i-n_j}. \nonumber
 \label{Ana_eqn_obc}
 \end{eqnarray}
The prefactor $\delta_n$ fixes the number of particles in the state to $N/q$ and is defined by
 \begin{eqnarray}
 \delta_n &=&
 \left\{\begin{array}{ll}
 \displaystyle
 1 & \mbox{for } \sum_i n_i= N/q,
 \\ [2ex]
 \displaystyle
 0 & \mbox{otherwise.}
 \end{array}\right.
 \end{eqnarray}
The other prefactor $\chi_n = (-1)^{\sum_j(j-1)n_j}$ is a phase factor, and the complex number $z_i$ denotes the position of the $i^{th}$ site in the two-dimensional plane. $n_i\in\{0,1\}$ denotes the particle occupation number at the $i^{th}$ site. The amplitude $\Psi(n_1, n_2, \ldots, n_N)$ can also be expressed as a CFT correlator.

\begin{figure*}
\includegraphics[width=17.8cm]{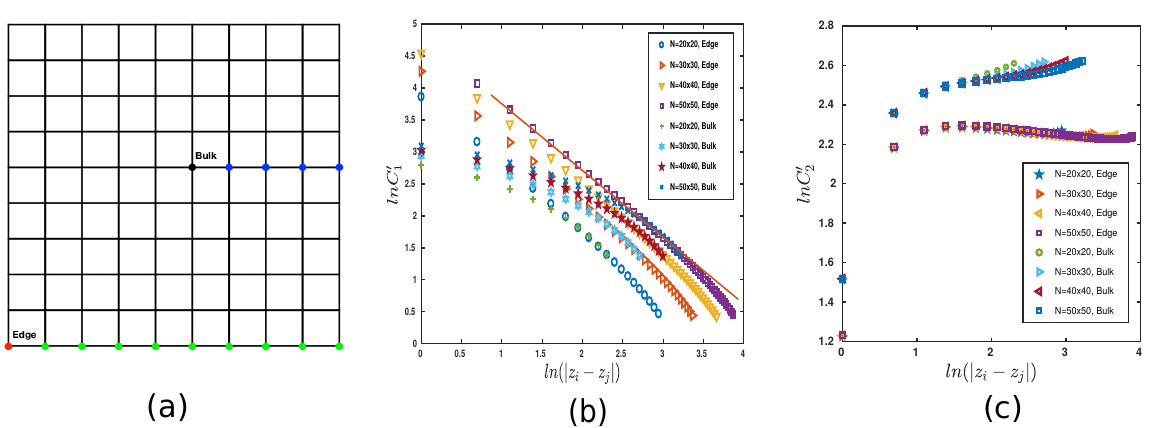}
\caption{(Color online) (a) To investigate the behavior of the two-body terms in the Hamiltonian, we consider the coupling strengths between the marked sites. The term `edge' refers to that the coupling strength is computed between the red site (site $i$) and each of the green sites (site $j$), and the term `bulk' refers to that the coupling strength is computed between the black site (site $i$) and each of the blue sites (site $j$). (b) Log-log plot showing the behavior of $C'_1 = \sum_{k (\ne i, \ne j)}(w^*_{ki}w_{kj} + w^*_{ji}w_{jk} + w^*_{ik}w_{ij})$ related to the coefficient of the $d^{\dagger}_id_j$ term with respect to distance $|z_i-z_j|$. The straight line (shown in red) has slope $-1$. (c) Log-log plot showing the behavior of $C'_2 = 2\sum_{k (\ne i,\ne j)}(w^*_{ij}w_{ik} + w^*_{ik}w_{ij} + w^*_{ji}w_{jk} + w^*_{jk}w_{ji})$ related to the coefficient of the $n_in_j$ term with respect to distance $|z_i-z_j|$.}
\label{coef}
\end{figure*}

\subsection{Exact Hamiltonian}
In the CFT construction of the Hamiltonian, null fields are used to derive a family of operators that annihilate the considered state. These operators are of increasing complexity, and for the lattice Laughlin states the three simplest are\cite{TuNJP}
\begin{eqnarray}
 \varUpsilon &=& \sum^N_{j=1} d_j,   \label{Up}  \\
 \Omega  &=& \sum^N_{j=1}(qn_j -1), \label{Om}  \\
 \Lambda_i &=& \sum_{j \ne i} w_{ij} [d_{j} - d_i(qn_{j} -1)]. \label{Lambda}
\end{eqnarray}
Here, $d_j$ is the hardcore bosonic (fermionic) particle annihilation operator at site $j$ for $q$ even (odd), and $w_{ij} = \frac{1}{z_i - z_j}$. If we construct the Hamiltonian from $\varUpsilon$ or $\Omega$, we do not obtain a unique ground state. We therefore construct the Hamiltonian from $\Lambda_i$ using Eq.\ \eqref{ham}. After multiplying out all the factors, we have
\begin{multline}
 H^{\textrm{Exact}} = \sum_{i \ne j} C_1(i,j) d^{\dagger}_i d_j + \sum_{i \ne j} C_2(i,j) n_i n_j \\
                + \sum_{i \ne j \ne k} C_3(i,j,k)d^{\dagger}_i d_jn_k + \sum_{i \ne j \ne k} C_4 (i,j,k) n_i n_jn_k  \\
                + \sum_{i} C_5(i)n_i, \label{hamexact}
\end{multline}
where the coefficients are given by
\begin{align}
& C_1(i,j) = 2w^*_{ij}w_{ij} + \sum_{k (\ne i, \ne j)}(w^*_{ki}w_{kj} + w^*_{ji}w_{jk} + w^*_{ik}w_{ij}), \nonumber\\
& C_2(i,j) = (q^2-2q)w^*_{ij}w_{ij} - q\sum_{k (\ne i,\ne j)}(w^*_{ij}w_{ik} + w^*_{ik}w_{ij}), \nonumber \\
& C_3(i,j,k) = -q(w^*_{ji}w_{jk} + w^*_{ik}w_{ij}), \nonumber \\
& C_4(i,j,k) = q^2 w^*_{ik}w_{ij}, \nonumber \\
& C_5(i) = \sum_{j(\ne i)} (w^*_{ji}w_{ji} + w^*_{ij}w_{ij}) + \sum_{j,k (\ne i)}w^*_{ij}w_{ik}. \nonumber
\end{align}
This Hamiltonian is not SU(2) invariant. It conserves the number of particles, and we shall assume throughout that the number of particles is fixed to $N/q$. The state \eqref{state} is the ground state of this Hamiltonian for any choice of lattice in the plane (i.e.\ for any choice of $z_j$).

\subsection{Behavior of the coefficients of the Hamiltonian}

There are five different types of operators in the Hamiltonian: a one-body term, two two-body terms, and two three-body terms with different coupling coefficients. The one-body term is local, since each term acts only on a single site. The coefficients $C_3$ and $C_4$ of the three-body terms do not involve any summation over indices. They hence depend only on the relative positions of the involved lattice sites, and their behavior is the same everywhere on the lattice. The $C_3$ coefficient has two terms both of which have power law decay, while the $C_4$ coefficient has one term which has a power law decay with distance. It is also seen that the decaying behavior of the three-body coefficients is independent of the system size.

The coefficients of the two-body terms do, however, have a more complicated behavior. $C_1$ and $C_2$ both contain two terms. The first decays as the square of the distance between the sites, and the second is a sum. Due to these sums, the coefficients depend on the positions of all the lattice sites. We demonstrate the behavior explicitly for the case of a square lattice in Fig.\ \ref{coef}. In this figure, we plot only the term containing the sum, and for $C_2$ we plot the sum of the contributions from $n_in_j$ and $n_jn_i$, since these two operators are the same. We consider two cases. In the first case, we fix $i$ to be an edge lattice point and then let $j$ run along the edge. In the second, we fix $i$ to be a lattice point in the middle of the lattice and then let $j$ run towards the edge. The two cases are illustrated in Fig.\ \ref{coef}(a).

From the results in Fig.\ \ref{coef}(b) and \ref{coef}(c), we make the following observations. First, the coefficients can be large even at distances comparable to the size of the lattice. Second, the behavior of the coefficients with distance depends on where the sites are on the lattice. Third, there is, in general, not a well-defined thermodynamic limit. In particular, this means that if we truncate the Hamiltonian directly, the strengths of the local terms in the bulk will not only depend on the local lattice geometry, but on the size and shape of the whole lattice. This problem does not arise when truncating the $\Lambda_i$ operators instead.

\section{Local model}\label{sec:local}

\begin{figure}
\includegraphics[width = 0.9\columnwidth]{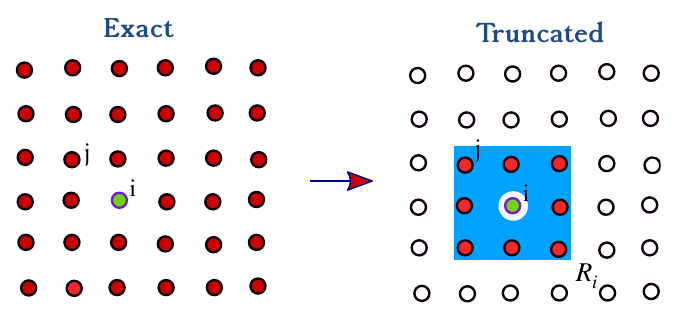}
\includegraphics[width = 0.9\columnwidth]{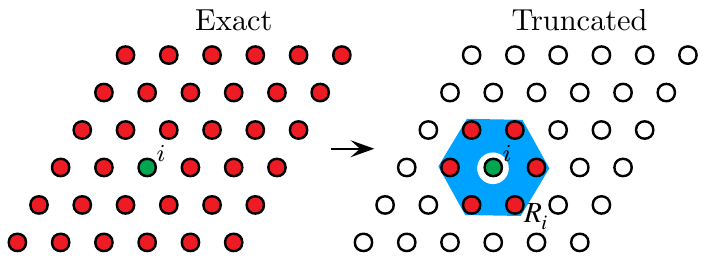}\caption{(Color online) Illustration of the truncation procedure of the $\Lambda_i$ operator for a square lattice (top panel) and a triangular lattice (bottom panel). The left part of the diagram shows the exact $\Lambda_i$ operator, which contains terms involving the $i^{th}$ lattice site (shown in green) and any of the other lattice sites (shown in red). The right part of the diagram displays the truncated form of the $\Lambda_i$ operator, which only contains terms involving the $i^{th}$ lattice site (shown in green) and any of the sites within a local region (shown in red with a blue background).}
\label{lamcut}
\end{figure}

The main goal of the present work is to investigate a general way to obtain a local Hamiltonian starting from the exact model. As discussed earlier, we start by truncating the operator $\Lambda_i$ to obtain the local operator
\begin{eqnarray}
 \Lambda^{\textrm{(L)}}_i = \sum\limits_{\substack{j \in R_i}} w_{ij} \left[d_{j} - d_i(qn_{j} -1)\right].
\end{eqnarray}
Here, the sum is taken over the sites within some local region around the $i^{th}$ site as illustrated in Fig.\ \ref{lamcut}. A natural choice is to let the local region contain all sites that are at most a distance $r$ away from the $i^{th}$ site. In that case, $R_i=\{j|\, 0 <|z_i - z_j| \le r\}$. The truncated Hamiltonian is then constructed as
\begin{eqnarray}
 H^{\textrm{Local}} = \sum^N_{i=1} \Lambda^{\textrm{(L)}\dagger}_i \Lambda^{\textrm{(L)}}_i.
\end{eqnarray}
The exact model is defined for open boundary conditions. This means that all the $\Lambda^{\textrm{(L)}}_i$ operators in the bulk have the same form, but the $\Lambda^{\textrm{(L)}}_i$ operators on the edge are different, because there are fewer neighbors. It is, however, straightforward to modify the truncated model to have periodic boundary conditions. We just need to use the bulk form of $\Lambda^{\textrm{(L)}}_i$ for all lattice sites and let the lattice wrap around the torus. Periodic boundary conditions reduce the finite size effects on the relatively small lattices we can investigate with exact diagonalization. In the following, we consider the truncated model for both open and periodic boundary conditions on square and triangular lattices.
\begin{figure}
\includegraphics[width = 0.9\columnwidth]{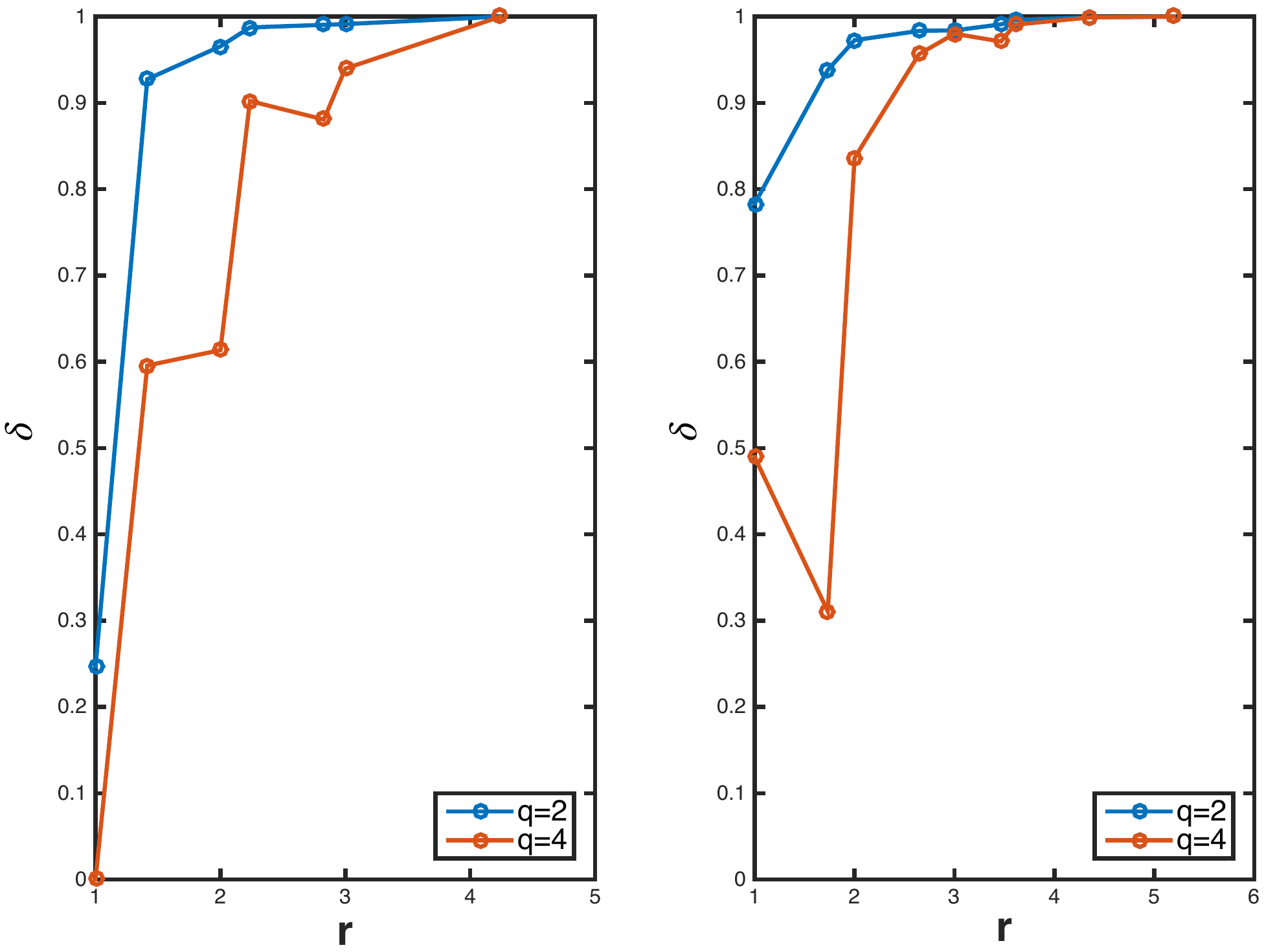}
\caption{(Color online) Overlap $\delta$  versus truncation radius $r$ computed for the case of a $4\times4$ square (left) or triangular (right) lattice with an open boundary.}
\label{obc}
\end{figure}

\section{Overlap of the eigenstates}\label{sec:gsprop}

In order to quantify how well the constructed local models reproduce the exact states, we compare the eigenstates of $H^{\textrm{Local}}$ computed from exact diagonalization to the exact states. On the plane, we define the overlap between the ground state $|\Psi_{\textrm{Local}}\rangle$ and the exact state \eqref{state} to be
\begin{equation}\label{deltapl}
\delta=|\langle \Psi_{\textrm{Local}}|{\Psi}^{\textrm{P}}_{\textrm{Exact}} \rangle|^2.
\end{equation}
In Fig.\ \ref{obc}, we plot the overlap $\delta$ as a function of truncation radius $r$ for the half ($q=2$) and quarter ($q=4$) filled cases on a $4\times4$ square and triangular lattice. For the half filled case, a truncation radius of $\sqrt{2}$ on the square lattice and $\sqrt{3}$ on the triangular lattice is enough to get an overlap higher than 0.9. For the quarter filled case, a truncation radius of $\sqrt{5}$ on the square lattice and $\sqrt{7}$ on the triangular lattice is needed to get an overlap higher than 0.9. Note that the number of particles per site is a factor of two smaller for $q=4$ than for $q=2$, so the same number of particles will take part in the local interactions if the truncation radius is about a factor of $\sqrt{2}$ larger for $q=4$. The behavior of the curves in Fig.\ \ref{obc} further suggests that the needed truncation radius to some extent depends on the shape of the local region.

\begin{figure}
\includegraphics[width = 0.9\columnwidth]{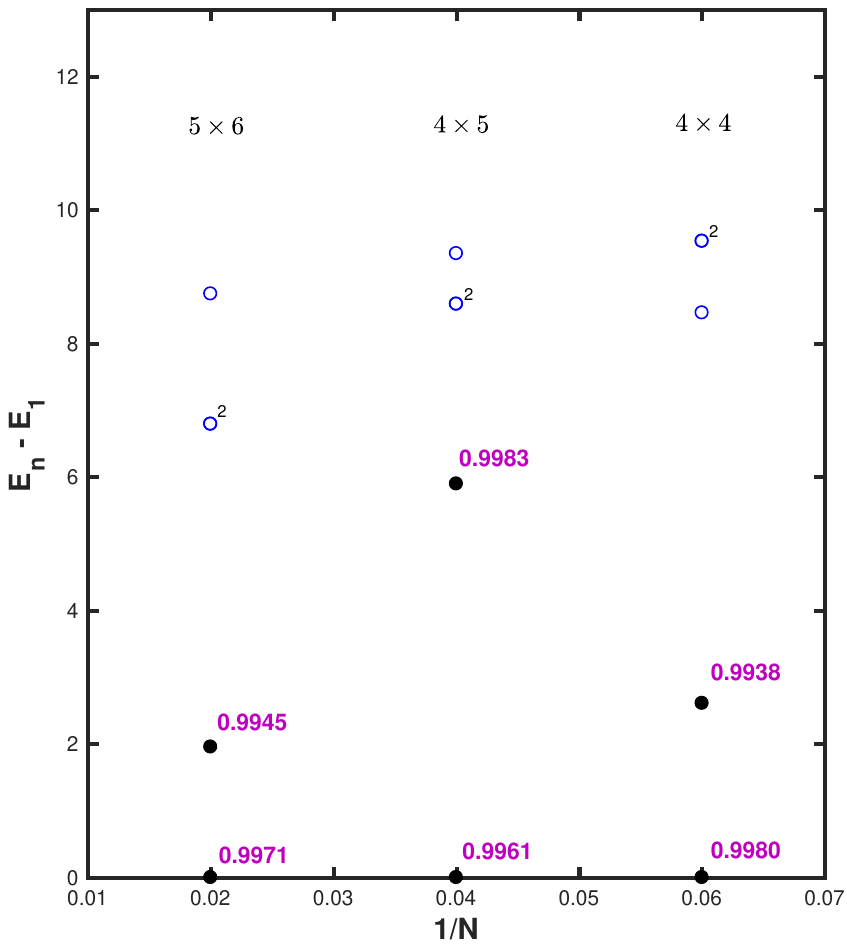}
\includegraphics[width = 0.9\columnwidth]{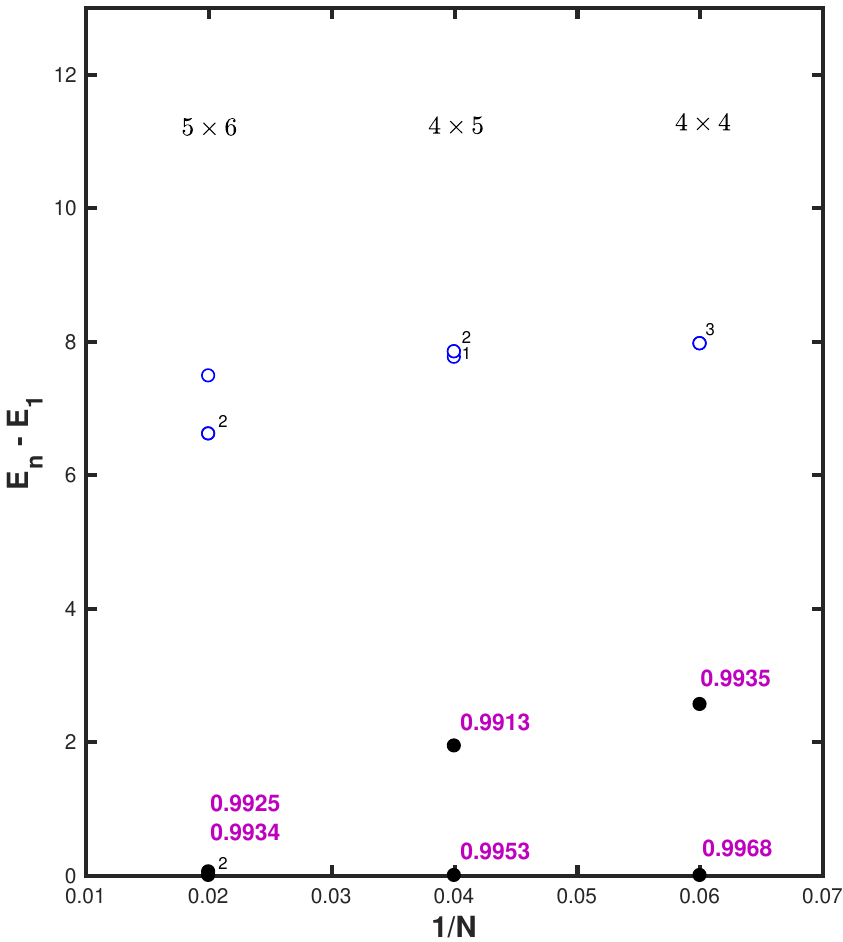}
\caption{(Color online) Energy spectrum for different lattice sizes ($N=L_x \times L_y$) on the torus for $q=2$ on a square lattice with $r=\sqrt{2}$ (top) and a triangular lattice with $r=1$ (bottom). The energy values painted black have a high overlap per site with the exact analytical states (the overlap per site is written in purple close to the eigenstate). The black integers give the degeneracies when different from 1.}
\label{gapq2}
\end{figure}

\begin{figure}
\includegraphics[width = 0.9\columnwidth]{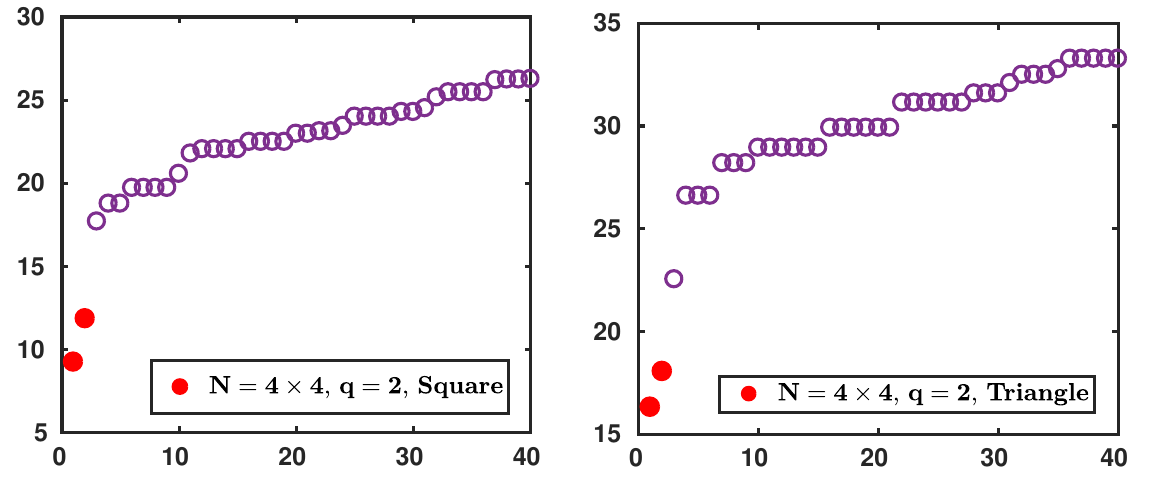}
\caption{(Color online) The first $40$ energy eigenvalues computed for $q=2$ on a $4\times 4$ square lattice with $r=\sqrt{2}$ (left) and a $4\times 4$ triangular lattice with $r=1$ (right). Only the states that have the same momentum quantum numbers as the analytical states can have a nonzero overlap, and the red points mark the $q$ lowest of these states.}
\label{energy}
\end{figure}

We now consider the case of periodic boundary conditions. Because of the topological order, there are $q$ states in total on the torus, which are given in the appendix. We therefore define the overlap between the $j^{th}$ eigenstate $|E_j\rangle$ of $H^{\textrm{Local}}$ and the exact state to be
\begin{equation}\label{delta}
\Delta =\sum_{l=0}^{q-1} |\langle E_j |\tilde{\Psi}^{\textrm{T},l}_{\textrm{Exact}} \rangle|^2,
\end{equation}
where $|\tilde{\Psi}^{\textrm{T},l}_{\textrm{Exact}} \rangle$ are the states in \eqref{psiT} after Gram-Schmidt orthonormalization. Due to the exponential growth of the Hilbert space dimension with system size, we expect the overlap to show an exponential decay with system size. This motivates us to also consider the overlap per site $\Delta^{1/N}$, which is expected to be roughly independent of system size for large enough systems.

Let us first investigate the model with $q=2$. The low energy part of the spectrum of the truncated Hamiltonian for different lattice sizes for both square and triangular lattices is shown in Fig.\ \ref{gapq2}, where we also provide the overlaps per site for the two lowest energy eigenstates. The truncation radius is $r=\sqrt{2}$ for the square lattice and $r=1$ for the triangular lattice. Figure \ref{energy} gives a more detailed view with more eigenstates for the $4\times 4$ lattices. The overlaps per site for the two lowest energy states are higher than $0.99$ for all the cases, which shows that the wavefunctions are close to the analytical states derived from CFT. Already for the $4\times 4$ lattice, the two lowest energy states are separated by a gap from the rest of the spectrum, which shows the twofold degeneracy on the torus, although not perfectly for this very small lattice size. The ratio $(E_2-E_1)/(E_3-E_1)$ is smaller for the $5\times 6$ lattice than for the $4\times 4$ lattice, and this suggests that a gap will be present in the thermodynamic limit.

\begin{figure}
\includegraphics[width = 0.7\columnwidth]{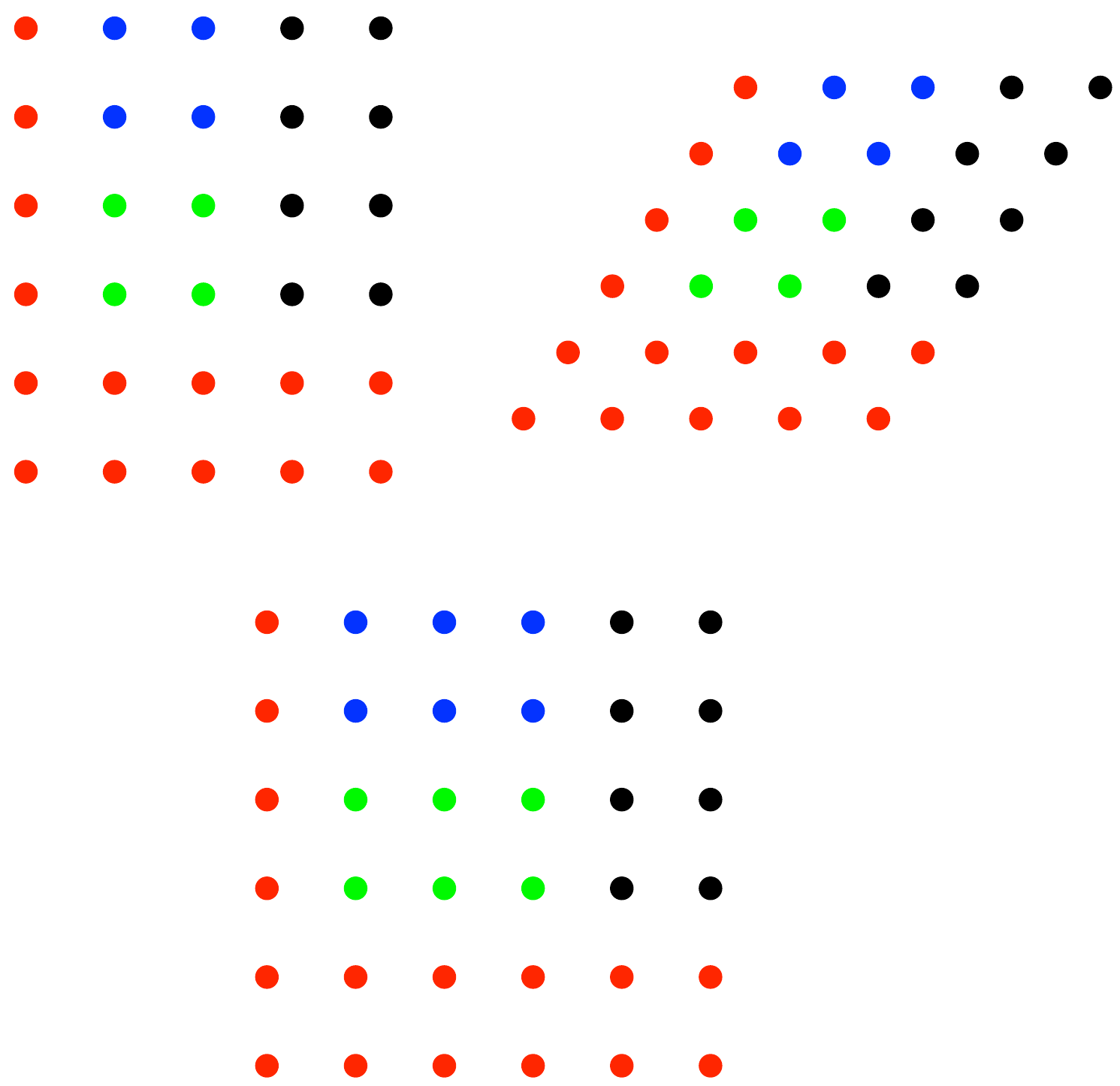}
\caption{(Color online) The division of the lattices used for computing the topological entanglement entropy. The different regions are marked with colors. The two upper plots are the $5\times 6$ lattices considered for $q=2$ and the lower plot is the $6\times 6$ lattice considered for $q=4$.}
\label{TEEdivision}
\end{figure}

As a further test of the topological nature of the ground states of the local models, we compute\cite{levin} the topological entanglement entropy $\gamma$ for the states on the $5\times 6$ lattice. The division of the lattice used for the computation is shown in Fig.\ \ref{TEEdivision}. We do not expect to get very accurate results for this lattice size, since the condition that all regions should be much larger than the correlation length is not entirely fulfilled. The exact lattice Laughlin state at filling fraction $1/q$ has $\gamma=\ln(q)/2$, which is $\gamma\approx0.347$ for $q=2$. For the square lattice, we obtain $\gamma=0.3216$ for the lowest energy eigenstate and $\gamma=0.3839$ for the second lowest energy eigenstate, and for the triangular lattice, we obtain $\gamma=0.3278$ for the lowest energy eigenstate and $\gamma=0.3412$ for the second lowest energy eigenstate. These values are close to the exact values and support the statement that the ground states are in the same topological phase as the half filled Laughlin states.

\begin{figure}
\includegraphics[width = 0.9\columnwidth]{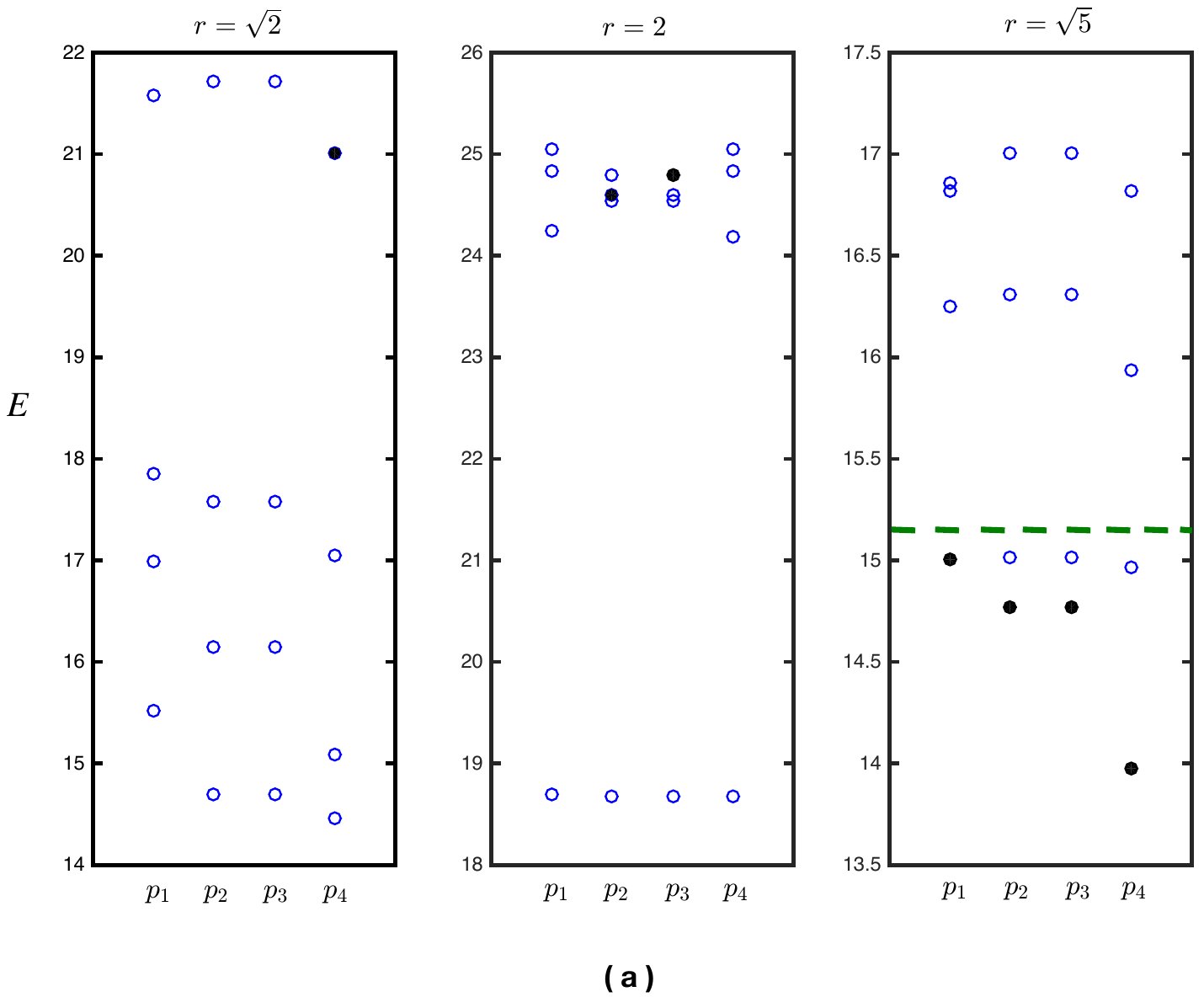}
\includegraphics[width = 0.9\columnwidth]{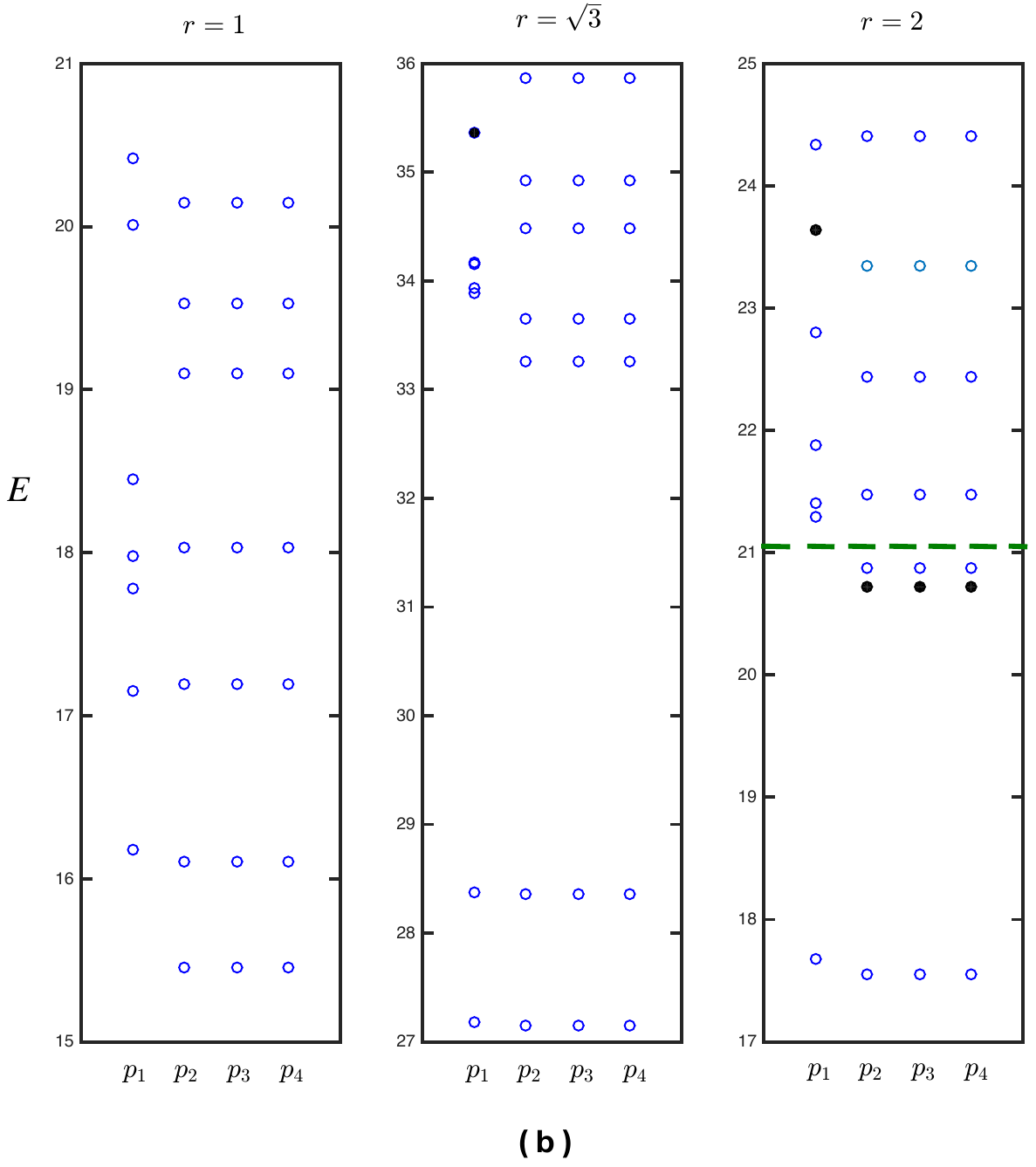}
\caption{(Color online) Energy spectrum in the relevant momentum sectors $p_i=(p_x,p_y)$ for different truncation radii computed on a $6\times6$ square (top) or triangular (bottom) lattice on a torus. Here $p_x$ and $p_y$ correspond to the momentum quantum numbers in the two periodic directions, respectively. The energy values painted black correspond to the eigenstates that have high overlap per site ($\Delta^{\frac{1}{N}}> 0.98$) with the exact analytical states. The green dashed line indicates the lowest energy present in the momentum sectors not shown.}
\label{gapq4}
\end{figure}

We next investigate the model with $q=4$. We saw above that this case requires a larger radius of truncation. Testing the local models on small lattices with open boundaries for different truncation radii is possible, but on the torus the local regions will wrap around the torus and overlap with themselves if they are chosen too large. For a given lattice size, this puts a restriction on the largest truncation radius that can be studied numerically. We hence study the spectra for $6\times6$ square and triangular lattices for different truncation radii as shown in Fig.\ \ref{gapq4}. The plots show the energy spectrum in the momentum sectors where the analytic states of interest lie. We observe that the overlaps for the ground states are small for small truncation radii as expected. The trend in the plots, however, indicates that the states with significant overlap with the exact analytic states climb down the energy spectrum as the truncation radius is increased. For the square lattice and $r=\sqrt{5}$, there are four states with high overlap among the low energy states, but there is not a clear gap. We are not able numerically to study large enough systems to judge whether a gap will be present in the thermodynamic limit or the system will be in a different phase. It is interesting to note, however, that both for the $4\times 4$ lattice with open boundary conditions studied in Fig.\ \ref{obc} and for the $6\times 6$ lattice with periodic boundary conditions, high overlap of the ground state is first observed for $r=\sqrt{5}$. This may suggest that $r=\sqrt{5}$ is also sufficient in the thermodynamic limits, but it would be necessary to investigate larger lattices to draw conclusions. For the local model at $r=\sqrt{5}$ on a $6\times6$ square lattice, we find $\gamma=0.7088,0.6231,0.6894,0.6724$ for the four lowest states that have significant overlap with the exact analytic states (the division used is shown in Fig.\ \ref{TEEdivision}). This is close to the expected value $\gamma=\ln(4)/2\approx 0.693$ and shows that the states with highest overlap with the exact analytic states are in the right topological phase.

For the triangular lattice, we also observe the trend that states with significant overlap are climbing down the spectrum as the truncation radius is increased. The results in Fig.\ \ref{obc} suggest that a truncation radius of at least $r=\sqrt{7}$ is needed to get high overlaps, which is consistent with the results in Fig.\ \ref{gapq4}. For the $6\times 6$ triangular lattice on the torus, we can, however, not study the case $r=\sqrt{7}$, since the largest allowed truncation radius is $r=2$.

\section{Conclusion}\label{sec:conclusion}

We have investigated a general procedure to truncate lattice Hamiltonians derived from CFT null fields. Each of the null fields leads to an operator $\Lambda_i$ that annihilates the state, and the approach is to truncate this operator to a local form instead of truncating the Hamiltonian itself.

Truncating the $\Lambda_i$ operator has a number of advantages compared to truncating the Hamiltonian directly. First, the terms in the operator depend only on the relative coordinates of the involved sites, and not on the rest of the lattice. The local terms are hence unaltered, when we approach the thermodynamic limit or in other ways modify the rest of the lattice. On the contrary, some of the terms in the Hamiltonian depend on the whole lattice, and the local terms resulting from truncating the Hamiltonian directly will depend on the choice and size of the whole lattice. Second, it is clear how to construct models with open or periodic boundary conditions as desired, while truncating the Hamiltonian directly only gives models with open boundary conditions, unless one manually changes the local terms and optimizes the coupling strengths numerically. The fact that numerical optimization is not needed means that we can construct the local Hamiltonian even if the local regions are not small compared to the lattice sizes that can be investigated with exact diagonalization.

We have applied the truncation procedure to Hamiltonians with half and quarter filled Laughlin type ground states on square and triangular lattices on the plane and on the torus. For the $q=2$ local model, we find that a truncation radius of $r=\sqrt{2}$ on the square lattice and $r=1$ on the triangular lattice is enough to obtain overlaps per site with the exact states that are larger than $0.99$ for all the investigated cases, and to obtain an approximate twofold degeneracy on the torus. This suggests that the truncated model is in the same topological phase as the exact model. We also find that the topological entanglement entropy is close to the expected value.

For the $q=4$ local model, we infer first from computations assuming open boundary conditions that a larger truncation radius is needed for the ground state of the local model to be close to the exact analytical state. We see a similar trend on the torus. Computations for the square lattice for open boundaries and on the torus seem to suggest that a truncation radius $r=\sqrt{5}$ is needed to stabilize the corresponding $q=4$ Laughlin state. The systems we can investigate with exact diagonalization are, however, too small to judge if the system is, indeed, topological in the thermodynamic limit. We have also checked that the states have the right topological entanglement entropy. For the case of the triangular lattice, the computations for open boundary conditions indicate that at least a truncation radius of $r=\sqrt{7}$ is needed for the ground state of the local model to be sufficiently close to the exact analytical state. However, numerical restrictions do not allow us to test this on larger lattice sizes on the torus.

An SU(2) invariant local Hamiltonian was constructed for the $q=2$ state on the square lattice and the kagome lattice in \onlinecite{natureanne} by truncating the Hamiltonian directly and numerically optimizing the coefficients. The same procedure does, however, fail to produce a simple Hamiltonian on the triangular lattice. In contrast, the approach of truncating the $\Lambda_i$ operator gives good overlap values on both the square and the triangular lattice without optimization. Also, for the Hamiltonian considered in this work there is no SU(2) symmetry to simplify the problem.

Constructing parent Hamiltonians using tools from CFT provides an alternative route to obtain FQH physics in lattice systems, compared to fractional Chern insulators. One advantage is that the Hamiltonians by construction have states with a particular topology as ground states. Local Hamiltonians are easier to work with both experimentally and theoretically, and it is therefore important to find suitable truncation schemes. In the future, it would be interesting to test how large the local regions need to be to give good overlaps for other FQH models derived from CFT.

\begin{acknowledgments}
This work has in part been supported by the Villum Foundation. DKN would like to thank the Max Planck Institute for the Physics of Complex Systems for hospitality during visits to the institute.
\end{acknowledgments}

\appendix
\section{Exact state on the torus}\label{sec:psiT}
The wavefunction \eqref{state} can be expressed in terms of a CFT correlation function, and by evaluating this correlation function on the torus, it is possible to also obtain the states on the torus.\cite{Deshpande} They have the following form
\begin{multline}
|\Psi^{\textrm{T},l}_{\textrm{Exact}}\rangle=\\
\sum_{n_1,n_2,\ldots,n_N}
\Psi^{\textrm{T},l}_{\textrm{Exact}}(n_1, n_2, \ldots, n_N)
|n_1,n_2,\ldots,n_N\rangle \label{psiT}
\end{multline}
with
\begin{multline}
 \Psi^{\textrm{T},l}_{\textrm{Exact}}(n_1, n_2, \ldots, n_N) \propto\\
 \delta_n \chi_n \; \Theta
 \begin{bmatrix}
  l/q + a_A \\ b_A
 \end{bmatrix} \left( \sum^N_i \xi_i (qn_i-1), q\tau \right) \\
 \times\prod_{i<j} \left(\Theta
 \begin{bmatrix}
  1/2 \\ 1/2
 \end{bmatrix}(\xi_i-\xi_j, \tau)\right)^{qn_in_j - n_i -n_j}.
 \label{Ana_eqn}
\end{multline}
Here, $\delta_n$ fixes the number of particles in the state to $N/q$, and $\chi_n$ is the same phase factor as for the states in the plane. There are $q$ states labeled by $l$, where $l\in\{0, 1, \ldots, q-1\}$. The parameters $a_A$ and $b_A$ in general depend on the number of lattice sites and the value of $q$, but when $N$ and $q$ are both even, as for all the cases studied here, $a_A=b_A=0$. For the square lattice $\tau = i L_y/L_x$, and for the triangular lattice $\tau = e^{\frac{i \pi}{3}}L_y/L_x$. Here, $L_x$ is the number of unit cells along the real axis, and $L_y$ is the number of unit cells in the $e^{i\pi/2}$ direction ($e^{i\pi/3}$ direction) for the square (triangular) lattice. The rescaled coordinates are given by $\xi_{i} = z_i/L_x$, where we assume that the lattice constant is set to unity. The Riemann theta function is defined as
\begin{eqnarray}
 \Theta
\begin{bmatrix}
 a \\ b
\end{bmatrix} \left(\xi, \tau \right) = \sum_{n \in Z} e^{i\pi\tau(n+a)^2 + 2\pi i(n+a)(\xi+b)}.
\end{eqnarray}
The states in \eqref{psiT} are not necessarily orthogonal, and we use Gram-Schmidt orthonormalization to numerically obtain an orthonormal set of states spanning the same space.

\end{document}